\documentclass[twoside,a4paper,11pt]{sea10}
\usepackage{graphicx}
\usepackage[bookmarks=false]{hyperref}
\usepackage{movie15}
\usepackage{float}
\restylefloat{figure}
\newcommand{\Ha}{\ifmmode {\rm H}\alpha \else H$\alpha$\fi\xspace}
\newcommand{\Hb}{\ifmmode {\rm H}\beta \else H$\beta$\fi\xspace}
\newcommand{\Hg}{\ifmmode {\rm H}\gamma \else H$\gamma$\fi}
\newcommand{\Hd}{\ifmmode {\rm H}\delta \else H$\delta$\fi}
\newcommand{\Heps}{\ifmmode {\rm H}\epsilon \else H$\epsilon$\fi}

\newcommand{\Hii}{\ifmmode \rm{H}\,\textsc{ii} \else H~{\sc ii}\fi}
\newcommand{\Nii}{[N~{\sc ii}]$\lambda$6584}
\newcommand{\nii}{\ifmmode [\rm{N}\,\textsc{ii}] \else [N~{\sc ii}]\fi}
\newcommand{\ionni}{\ifmmode [\rm{N}\,\textsc{i}] \else [N~{\sc i}]\fi}

\newcommand{\oi}{\ifmmode [\rm{O}\,\textsc{i}] \else [O~{\sc i}]\fi}
\newcommand{\Oii}{[O~{\sc ii}]$\lambda$3727}
\newcommand{\neiii}{\ifmmode [\rm{Ne}\,\textsc{iii}] \else [Ne~{\sc iii}]\fi}
\newcommand{\nev}{\ifmmode [\rm{Ne}\,\textsc{v}] \else [Ne~{\sc v}]\fi}
\newcommand{\hei}{\ifmmode [\rm{He}\,\textsc{i}] \else [He~{\sc i}]\fi}
\newcommand{\Hei}{\ifmmode \rm{He}\,\textsc{i} \else He~{\sc i}\fi}
\newcommand{\heii}{\ifmmode [\rm{He}\,\textsc{ii}] \else [He~{\sc ii}]\fi}
\newcommand{\Heii}{\ifmmode \rm{He}\,\textsc{ii} \else He~{\sc ii}\fi}
\newcommand{\oii}{\ifmmode [\rm{O}\,\textsc{ii}] \else [O~{\sc ii}]\fi}
\newcommand{\Oiii}{[O~{\sc iii}]$\lambda$5007}

\newcommand{\oiii}{\ifmmode [\rm{O}\,\textsc{iii}] \else [O~{\sc iii}]\fi}

\newcommand{\sii}{\ifmmode [\rm{S}\,\textsc{ii}] \else [S~{\sc ii}]\fi}
\newcommand{\siii}{\ifmmode [\rm{S}\,\textsc{iii}] \else [S~{\sc iii}]\fi}
\newcommand{\feii}{\ifmmode [\rm{Fe}\,\textsc{ii}] \else [Fe~{\sc ii}]\fi}
\newcommand{\feiii}{\ifmmode [\rm{Fe}\,\textsc{iii}] \else [Fe~{\sc iii}]\fi}
\newcommand{\fevii}{\ifmmode [\rm{Fe}\,\textsc{vii}] \else [Fe~{\sc vii}]\fi}
\newcommand{\ariii}{\ifmmode [\rm{Ar}\,\textsc{iii}] \else [Ar~{\sc iii}]\fi}
\newcommand{\ariv}{\ifmmode [\rm{Ar}\,\textsc{iv}] \else [Ar~{\sc iv}]\fi}
\newcommand{\cliii}{\ifmmode [\rm{Cl}\,\textsc{iii}] \else [Cl~{\sc iii}]\fi}
\newcommand{\atflux}{\ifmmode \langle \log t_\star \rangle_L \else $\langle
\log t_\star \rangle_L$\fi}

\newcommand{\niiHa}{\ifmmode \log [ \rm{N}\,\textsc{ii} ]/ \rm{H}\alpha \else
$\log [ N~{\sc ii} ]/ \rm{H}_\alpha $\fi}

\begin{document}
\pagenumbering{arabic}
\pagestyle{myheadings}
\thispagestyle{empty}
\vspace*{0.2cm}
\begin{flushleft}
{\bf {\LARGE
%
Recovering physical properties from narrow-band photometry
%
}\\
\vspace*{1cm}
%
W. Schoenell$^{1,2}$,
R. Cid Fernandes$^{2}$,
N. Ben\'\i tez$^{1}$
and
N. Vale Asari$^{3,4}$
%
}\\
\vspace*{0.5cm}
%
$^{1}$
Instituto de Astrof\'\i sica de Andaluc\'\i a (CSIC) \\
$^{2}$
Departamento de F\'{\i}sica - CFM - Universidade Federal de Santa Catarina \\
$^{3}$
Institute of Astronomy, University of Cambridge\\
$^{4}$
CAPES Foundation, Ministry of Education of Brazil
%
\end{flushleft}
%
\markboth{
Recovering physical properties from narrow-band photometry
}{ 
%
Schoenell et al.
%
}
\thispagestyle{empty}
\vspace*{0.4cm}
\begin{minipage}[l]{0.09\textwidth}
\ 
\end{minipage}
\begin{minipage}[r]{0.9\textwidth}
\vspace{1cm}
\section*{Abstract}{\small
%

Our aim in this work is to answer, using simulated narrow-band photometry data,
the following general question: What can we learn about galaxies from these new
generation cosmological surveys? For instance, can we estimate stellar age and metallicity
distributions? Can we separate star-forming galaxies from AGN? Can we measure
emission lines, nebular abundances and extinction? With what precision?

To accomplish this, we selected a sample of about 300k galaxies with good S/N from the SDSS
and divided them in two groups: 200k objects and a template library of 100k. We corrected
the spectra to $z = 0$ and converted them to filter fluxes. Using a statistical
approach, we calculated a Probability Distribution Function (PDF) for each property of
each object and the library. Since we have the properties of all the data from the
{\sc starlight}-SDSS database, we could compare them with the results obtained
from summaries of the PDF (mean, median, etc).

Our results shows that we retrieve the weighted average of the log of the galaxy
age with a good error margin ($\sigma \approx 0.1 - 0.2$ dex), and similarly for
the physical properties such as mass-to-light ratio, mean stellar metallicity,
etc. Furthermore, our {\bf main result} is that {\bf we can derive emission line
intensities and ratios} with similar precision. This makes this method unique in
comparison to the other methods on the market to analyze photometry data and
shows that, from the point of view of galaxy studies, future photometric surveys
will be much more useful than anticipated.

%
\normalsize}
\end{minipage}
%
%
%
\section{Introduction \label{intro}}

This paper is essencially motivated due to the J-PAS project
http://j-pas.org, which in a near future will produce 8000 sq
degrees of imaging in 56 filterbands of 135\AA\ up to magnitude $i_{AB}
\approx 24$. The survey, besides producing a very high-quality
photometric redshift catalog, will provide highly informative data
important to other astronomical communities as galaxy evolution.

Here we describe a bayesian method to ``boost the spectral resolution'' of these
kind of photometric data. The idea, which will be described in more detail on
section \ref{sec:method}, is basically that if a high-resolution degrated
spectra (or a set of them) is similar to an observed j-spectra\footnote{We define a 
j-spectrum by the set of photometric measurements on J-PAS filter system.} then its
physical properties (and even emission lines) would be similar.

To test the efficiency of this method, we used the {\sc starlight}-SDSS 
\cite{Cid05} http://starlight.ufsc.br/ database as a sandbox.  We downloaded
the 926246 galaxies where there are measured physical properties and emission
lines.


\section{From SDSS to JPAS}
\label{sec:SDSS2JPAS}

Given a SDSS observed spectra, one can convert its energy distribution
$O_\lambda$ to an arbitrary observed photometric filter $l$ defined by its
transmission curve $T_{l,\lambda}$ using the simple conversion $J_l = \frac{\int
O_\lambda T_{l,\lambda} d\lambda}{\int T_{l,\lambda} d\lambda}$ and the error on
the filter $l$, $\sigma^2(J_l)$, with the relation $\sigma^2(J_l) = \Lambda_l
\langle  \sigma(n_\lambda)^2 t_{l,\lambda}^2 \rangle \Delta \lambda$ where
$\Lambda_l \equiv  N_{\lambda,l} \Delta \lambda$ is the effective filter size
and $n_\lambda$ is the spectral error in each point.

\begin{figure}
\center
	\includegraphics[width=8.3cm]{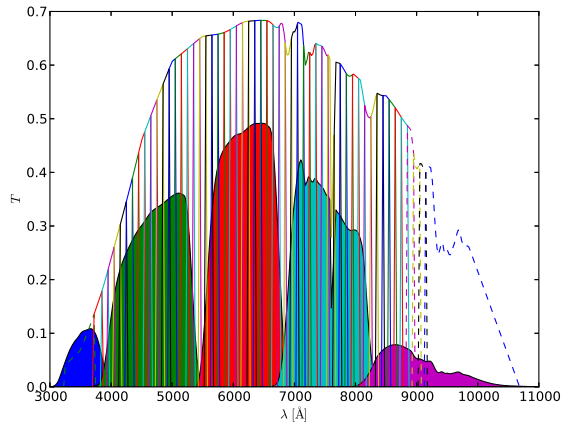}
	\caption{JPAS 56 filtersystem sensitivity curves. The 51 solid lines are the
	filters used on this work assuming mirror and CCD characteristics and an
	airmass of 1.3.
	We also plotted in solid fill the SDSS {\em u}, {\em g}, {\em r}, {\em i} and
	{\em z} filters in the same conditions.}
\label{fig:FilterSet}
\end{figure}

The filtersystem curves to JPAS considering an airmass of 1.3 and the expected
CCD plus Telescope efficiencies are shown on figure \ref{fig:FilterSet}. There
will be 56 filters, but due to the spectral coverage of SDSS we removed the
first one in the blue and the four last filters giving us an filtersystem of 51
filters plotted in solid lines. For comparsion, we plotted right below our
filtersystem the SDSS filterset {\em u}, {\em g}, {\em r}, {\em i} and {\em z}
sensitivities through the same air mass.


\begin{figure}
\center
    \includegraphics[width=8.3cm]{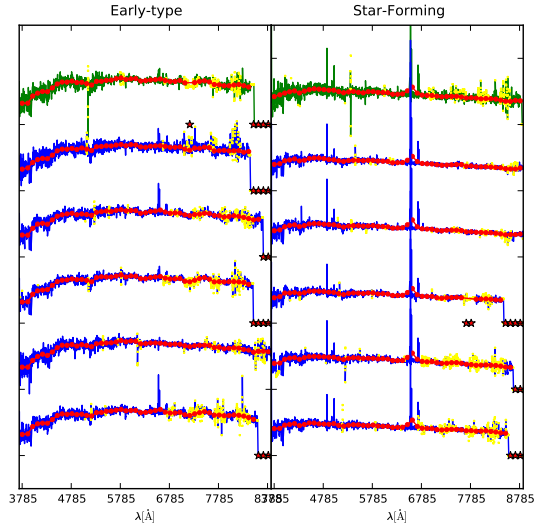}
    \caption{Two examples of j-spectrum and its 5 best matches based on $\chi^2$.
    At the top, in green, the two observed spectra examples: an early-type on  left panel
    and a star-forming galaxy on the the right. In blue, from top to bottom, the
    first five best matches. Their j-spectrum is represented by the connected red
    dots. Yellow points are flagged as bad pixels and were not considered.
    Stars represent the points were j-spectrum is not available.}
\label{fig:gal_examples}
\end{figure}

The main idea of this work is to obtain high precision properties from
low resolution data by taking a shortcut from what we measure from high
resolution spectra. We consider that if an pair object-template
j-spectra are similar, then their measured properties will be similar as
well. On fig \ref{fig:gal_examples} we show two examples of matching two
objects with their five best matches in respect to our template library
library\footnote{In this paper all the galaxies which belongs to the
comparsion sample we call library template galaxies.}. In green we
plotted the SDSS object and in blue its four best matches accordingly to
their $\chi^2$ calculated over their j-spectrum (plotted as red dots).
On the right panel we have an example of a star-forming
galaxy and on the left an early-type.

This idea is similar to the adopted by \cite{Gallazzi05} and
\cite{Kauffmann03a} and others, but with an important difference: Here
we do not compare an observed spectrum with a set of models but we
compare it to a set of another observed spectra. This not only permits
us to measure the standard physical properties such as mean age and
stellar masses that can be measured by ohter methodos, but it also
allows us to measure indirectly emission lines on data which we
evidently do not have enough resolution. This is the most relevant
advantage of this method.

\section{Method}
\label{sec:method}

To derive the properties $p$ listed down on table \ref{tab:Results}, we
calculated for 100k objects and two samples of comparsion their
likelihood functions ${\cal L}_{i,j} = e^{-f_L\frac{1}{2}\chi^2_{i,j}}$.
Where $\chi^2_{i,j} = \sum_l \frac{1}{N_{\rm good}} (O_{l,i} - a_{i,j}
B_{l,j})^2 w_{l,i,j}^2$ with a scaling factor $a_{i,j} = \frac{\sum_l
O_{l,i} B_{l,j} w_{l,i,j}^2}{\sum_l B_{l,j}^2 w_{l,i,j}^2}$ which is
determined by interacting the calculation of $a$ and $\chi^2$ until a
convergence critery of $\Delta a_{i,j} < 10^{-5}$ is accomplished. In
our simulations, this takes no more than four interactions. The $f_L$
term adjusts the width of the PDF. It can be adjusted to minimize
errors, but we will not treat it here.

The weighting used to the spectra was defined by $w^2_{l,i,j} = \left(
\frac{\langle O^2_{l,i} \rangle}{O^2_{l,i}} \right) \left(
\frac{1}{\sigma^2(O_{l,i}) + a^2_{i,j} \sigma^2(B_{l,i})} \right)$. This
was selected to have an unbiased weight in the amplitude of the spectra.
In other words, we would like to give the same importance to the parts
with high fluxes (e.g. emission lines) than that to that regions that
have lower ones (e.g. continuum).

With the likelihood for each object and base, we calculated as the output of our
method a PDF estimator. In our case, to estimate the output, we used the
likelihood-weighted average
$	\overline{p_i} = \frac{\sum_{j} p_j {\cal L}_{i,j}}{\sum_{j} {\cal L}_{i,j}} =
	p_{out,i}$

\begin{figure} 
	\center
    \includegraphics[width=8.3cm,clip=true]{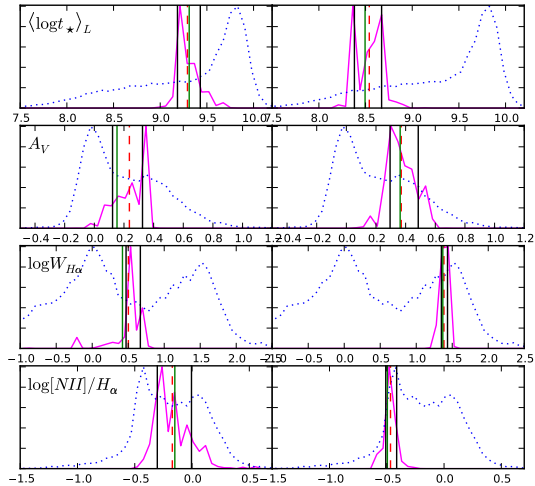}
    \caption{Examples of normalized PDF distributions to age, extinction, H$\alpha$\ 
    and [N~{\sc ii}]/H$\alpha$. The left and right boxes corresponds to the left and right
    observed galaxaies shown on fig. \ref{fig:gal_examples}. In blue dashed line
    we show the distribution of our base of templates, in magenta the likelihood distribution (or
    the posterior) and the horizontal lines represents in solid black the 16th
    and 84th percentiles, in red dashed the average and in solid green the value
    measured by {\sc starlight} directly on the spectrum.}
\label{fig:pdfs}
\end{figure}

\section{Sample selection}
\label{sec:samples}

Our main sample was retrieved from the {\sc starlight}-SDSS database based on
very wide criteria, trying to get all kinds of objects. Firstly, we separated
from the database the galaxies which were in the SDSS main galaxy sample \cite{YorkSDSS},
then we selected those who do not have any bad pixel in intervals of
31\AA\ centered on the emission lines H$\alpha$, H$\beta$, \Nii, \Oii\ and \Oiii\ to assure
that when we do not measure an emission line it is because it is too weak to
measure and not because of any observational error. A last selection in redshift
($0.01 \leq z \leq 0.11$) was applied in order to have spectral coverage on the
wavelength interval where JPAS will observe. Those criteria reduced the total
number of galaxies on our sample from 926246 to 299253 galaxies. All these
galaxies were corrected to the rest-frame. As a quick-look test, we will not
treat the redshift as a variable here and all results will be shown to $z = 0$.

We then converted all the observed spectra to J-spectra. In case of
problems on the SDSS spectra (e.g. bad pixels), we changed the observed
flux to the best fit from the {\sc starlight} in the cases where the
filters have less bad pixels than 50\% of the filter width. Otherwise,
the filter is flagged to be neglected.

Then, we divided the sample in two sub-samples: one with the galaxies with $S/N >
20$ (113821 galaxies) which we call mother library and other one with galaxies with
$S/N < 20$ (185432 galaxies) which we call object sample. All comparsions made
in this paper will be in respect of a set of objects and their PDFs calculated
over a given library which is a set of galaxies from the mother library. From the
mother library, we selected two samples of galaxies based on two independent
diagrams.

The first one, which we call CMD, is based on the physical analog to the
color-magnitude diagram to galaxies, the $\log\ M_\star$ -- $\langle \log\
t\rangle$ diagram.
We chopped this diagram in boxes of 0.1 dex, and on each box we got 10\% of the
objects distributed along the $A_V$ axis. This gives us 11952 galaxies on this
library.

The second library, which we call WHAN, is based on the WHAN diagnostic diagram
introduced by \cite{Cid10} and divide the galaxies basically between
star-forming, active nuclei and passive or retired galaxies. We did the same cut
on this diagram as we did on CMD library, but here we changed the $A_V$ to the
emission line ratio $\log \nii/{\rm H}alpha$. This gives us 27537 galaxies on this library.

\section{Results}
\label{sec:results}

To estimate our method's precision, we compared the properties derived
by our method (output) on the low-resolution j-spectra with the values
derived using {\sc starlight} on the high-resolution SDSS spectra
(input). So, to do this comparsion, we evaluate, for each property $p$
and object $i$, the $ \Delta p_i = p_{i,\rm output} - p_{i,\rm input}$.

This experiment shows the potential of the proposed method and
accomplish with our initial objective which is to test the precision of
galaxy properties with it. From table \ref{tab:Results}, we can resume
the precision of our method: It can measure physical properties like age
and extinction with typical precision of $0.2$ dex (or mag, in the case
of $A_V$) and emission lines with $0.3$ dex and our library selection
does not affected the final result, probably because we have a number of
templates which are in both libraries and/or beacause we have
oversampled libraries with $N_{\rm galaxies}$ of about 10k galaxies.

Once more, our main result is that {\bf we can measure emission lines without having 
sufficient spectral resolution to do it directly on our data}.

\begin{table}[ht]
\centering
\caption{Results} 
\begin{tabular}{ccccc} 
\hline\hline 
Property	&	$\overline{\Delta p}$	&	$\sigma ( \Delta p )$ (CMD)	&	$\overline{\Delta p}$	&	$\sigma ( \Delta p )$ (WHAN) \\ [0.5ex]   
\hline
$A_V$	&	0.023	&	0.106	&	0.023	&	0.101	\\	
$\langle \log t_\star \rangle_L$	&	-0.018	&	0.199	&	-0.013	&	0.192	\\	
$\langle \log Z_\star \rangle_L$	&	-0.021	&	0.144	&	-0.020	&	0.141	\\	
$ \log M / L_r$	&	-0.045	&	0.114	&	-0.040	&	0.110	\\	
$\log W_{ [ O_{II} ] }$	&	0.051	&	0.223	&	0.040	&	0.218	\\	
$\log W_{H\beta}$	&	0.024	&	0.145	&	0.030	&	0.143	\\	
$\log W_{ [ O_{III} ] }$	&	0.046	&	0.245	&	0.029	&	0.232	\\	
$\log W_{H\alpha}$	&	0.010	&	0.160	&	0.022	&	0.157	\\	
$\log W_{[ \rm {N} II ]}$	&	-0.028	&	0.159	&	-0.010	&	0.156	\\	
$\log [ N_{II} ] / H_\alpha$	&	-0.045	&	0.141	&	-0.044	&	0.146	\\	
$\log [ O_{III} ] / H_\beta$	&	0.026	&	0.250	&	-0.000	&	0.238	\\	
$\log H_\alpha / H_\beta$	&	-0.011	&	0.107	&	-0.010	&	0.114	\\	
$\log S_{II} / H_\alpha$	&	-0.006	&	0.172	&	-0.016	&	0.174	\\	
$\log [ O_{II} ] / H_\beta$	&	0.036	&	0.202	&	0.016	&	0.198	\\	
$\log [ O_{III} ]/ [ N_{II} ]$	&	0.075	&	0.265	&	0.044	&	0.252	\\	
\end{tabular} 
\label{tab:Results}
\end{table}

%
%
\small  
%
\section*{Acknowledgments}   
%

We thank financial support from CNPq, FAPESP, Instituto de Astronom\'\i a de
Andaluc\'\i a and the CNPq's Instituto Nacional de Ci\^encia e Tecnologia -
Astrof\'\i sica. NVA has been supported by CAPES (proc. no. 6382-10-0). WS has been
supported by Spanish Ministerio de Econom\'\i a y Competitividad (grant
AYA2010-22111-C03-01) and Brazilian INCT-A. The SEAGal Team wishes to thank all researchers
involved in the Sloan Digital Sky Survey for their dedication to a project which has made
the present work possible.

%

%
\end{document}